# Labeling elemental detection sensitivities in part per billion range using conventional geometry synchrotron assisted EDXRF measurements.

Md. Akhlak Alam[1, 2], M. K. Tiwari[1, 2,]*, Ayushi Trivedi[1], Ajay Khooha[1], A. K. Singh[1]

[1] Synchrotrons Utilization Section, Raja Ramanna Centre for Advanced Technology, Indore - 452013, India.

[2] Homi Bhabha National Institute, Anushaktinagar, Mumbai - 400094, India.

*Corresponding author: mktiwari@rrcat.gov.in

**Abstract**

Energy dispersive X-ray fluorescence (EDXRF) is a widely used non-destructive technique for micro and trace multi-element analysis of materials. Conventional trials show that using laboratory assisted EDXRF measurements, one can obtain elemental detection limits in the range of µg/g to sub-µg/g level. In the present work a quantitative approach has been followed in attempting to explore how is it possible to obtain elemental detection limit in the range of ng/g by using simple EDXRF excitation (45°- 45° geometry) instead of using total reflection X-ray fluorescence (TXRF) technique, which renders relatively superior detection limits for different elements. In order to accomplish this, we recorded fluorescence spectrum from a standard reference sample (ICP-IV) in similar experimental conditions. The results show that using a very small quantity of sample on top of a thin kapton foil with a thickness ranging between 25-50 µm, as a sample carrier, the EDXRF technique may offer comparable elemental detection limits in contrast to TXRF technique.

# INTRODUCTION

X-ray fluorescence (XRF) spectrometry is one of the commonly used analytical technique which gives element specific information for a material (in solid or liquid phase) in a non-destructive approach. The technique is frequently employed for multi-element micro and trace analysis of ultra-pure chemical, biological, geological and environmental materials [1-6]. Total reflection X-ray fluorescence (TXRF) is another variant of energy dispersive XRF (EDXRF) method wherein main difference appears in its excitation geometry. In TXRF, the primary X-ray beam impinges on a sample carrier at a grazing incidence angle smaller than the critical angle of the sample support. Small amount of sample deposited in the suspension form on top of the polished substrate is excited by both incident as well as totally reflected X-ray beams. Owing to small sample mass used in the TXRF, generally matrix effects (absorption and enhancement) are neglected during quantitative analysis as compared to that of the EDXRF technique [7].

In general, using conventional EDXRF approach (45°- 45° excitation and detection geometry) it is possible to obtain detection limit (DL) values of the order of few microgram levels or below, depending on the nature of different excitation sources [8, 9]. On the other hand, TXRF excitation geometry provides detection limit values of the order of few nano grams or even below. Improved elemental detection limit values mainly arises due to lower background ascribed by minimal penetration of primary X-ray beam inside the sample support. Almost all elements with atomic number $Z > 11$ can easily be detected with the help of a laboratory X-ray source assisted TXRF instrument in ambient air environment. Attempts to evolve practical and applied approaches that can improve detection limits of the conventional EDXRF technique comprising of 45°-45° experimental geometry, play an important role for its wide range of applications [10-13].

In this article, we report a simple approach that allows one to obtain improved detection limits for different elements present in an analyte using conventional EDXRF excitation geometry. We have shown that by employing kapton foil as sample carrier in the thickness range of 25-50 μm, and depositing very small mass of the specimen on top of it, improved detection limits can be realized. The enhancement of fluorescence signal to background ratio mainly arises due to reduced spectral background intensity originating from thin kapton foil as well as additional excitation of an analyte element caused by Compton and Rayleigh scattered X-rays. The present work aims to find a modest low-cost method for routine ultra-trace element analysis using simple EDXRF excitation geometry that can compete with the capabilities of TXRF method. We further realized that kapton foil may be considered as a convenient economic option to be used as sample carrier for EDXRF analysis of solid and liquid specimens.

**EXPERIMENTAL**

Both EDXRF and TXRF measurements were performed at BL-16 beamline of Indus-2 synchrotron facility [14] in order to compare and correlate detection limits observed with the two experimental approaches. In order to investigate detection sensitivities of different elements over the full spectral background we used a multielement reference material instead of using a single element standard. For this a liquid multi-element standard reference sample (ICP-IV) [15], comprising of 23 elements ( Ag, Al, B, Ba, Bi, Ca, Cd, Co, Cr, Cu, Fe, Ga, In, K, Li, Mg, Mn, Na, Ni,

Pb, Sr, Tl, Zn) with concentrations of each element ~ 1000 μg/ml was used. First, ICP-IV sample was diluted up to concentration value of ~ 4.5 μg/ml by mixing it in ultra-pure demineralised water and then ~ 20 micro-litre volume of diluted specimen was deposited (providing ~ 90 ng mass of the analyte) on top of kapton foils of different thickness values ranging from 12 to 100 μm. In the case of TXRF analysis, same amount (~ 90 ng) of the ICP-IV specimen was deposited on a highly polished quartz glass substrate. Fluorescence measurements were carried at incident X-ray energy of 20 keV, monochromatized using a Si (111) double crystal monochromator (DCM). During EDXRF measurements an X-ray beam of size of ~2 (h) × 2 (v) mm$^2$, obtained from a cross slit attachment was allowed to excite the analyte residue deposited on top of the kapton foil, ensuring full illumination of deposited sample. On the other hand in case of TXRF measurements, X-ray beam of the size ~10 (h) × 0.1 (v) mm$^2$ was utilized. Fluorescence X-rays emitted from the sample residue were recorded using a silicon drift detector (SDD, Amptek, USA) comprising of an active area ~ 25 mm$^2$. The SDD system was placed at a distance of ~35 mm from the centre of sample (in both EDXRF and TXRF experimental geometries) and fluorescence spectra were acquired for a detector live time of 100 seconds in each case. Figs. 1(a) and (b) respectively show the schematic layouts of the EDXRF and TXRF measurement geometries. In both the cases, SDD fluorescence detector was placed in the plane parallel to S-polarization plane of synchrotron X-ray beam. In Fig. (2), we have depicted actual photograph of the experimental setup used for the EDXRF measurements at BL-16 beamline of Indus-2 synchrotron facility.

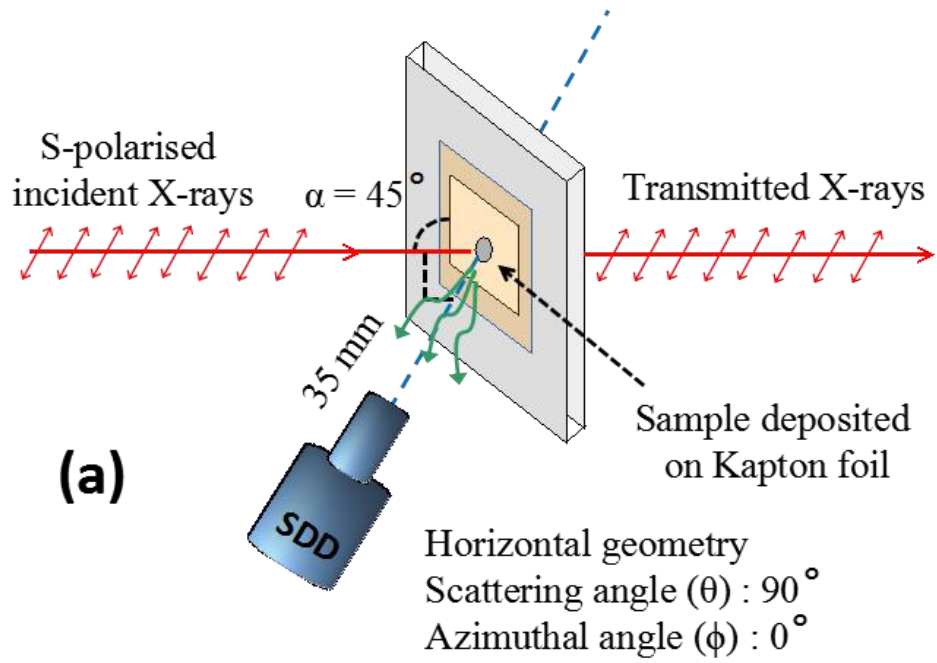

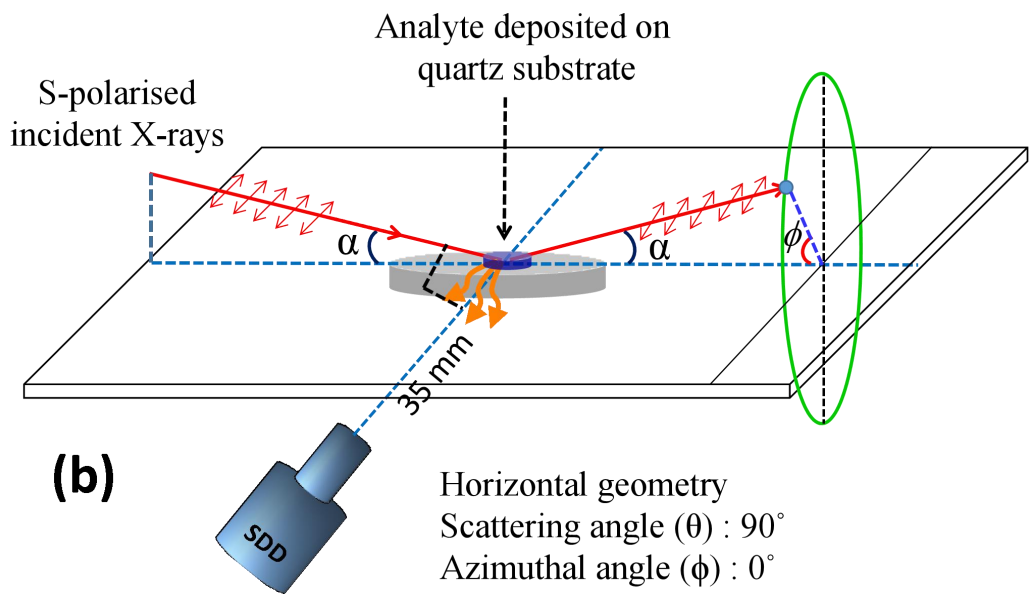

**Fig. 1.** Schematics illustration showing the experimental geometries (not to scale) used for (a) EDXRF and (b) TXRF measurements.

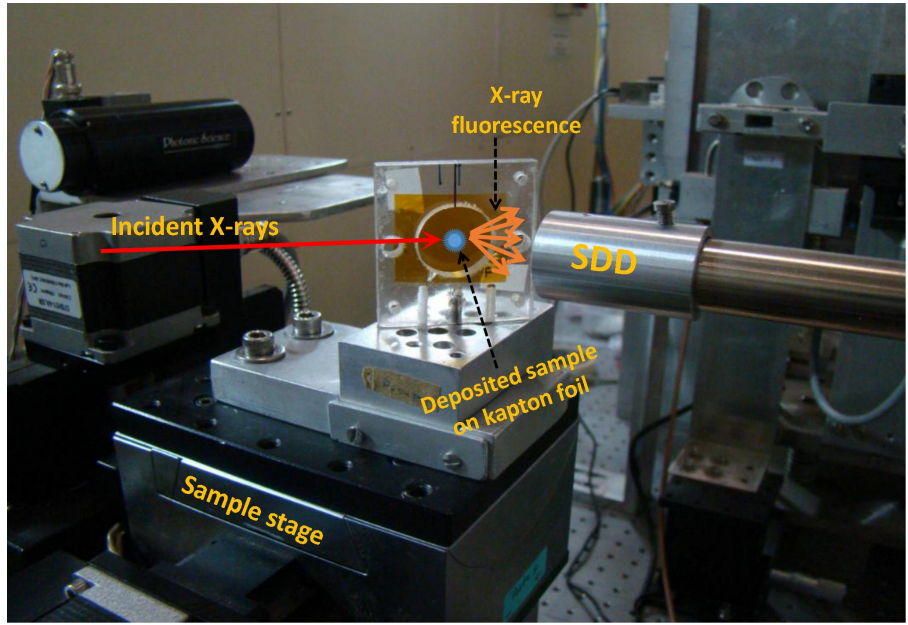

**Fig. 2.** Photograph of the experimental set-up used for the EDXRF measurements at BL-16 beamline of Indus-2 synchrotron facility.

## RESULTS AND DISCUSSION

**Theoretical calculations:**

X-ray interacts with matter either by scattering or absorption processes. X-ray fluorescence photons are generated from a material as a result of interaction of primary X-ray photons with core shell orbital electrons of an absorbing atom. The spectral background during EDXRF measurement, on the other hand, is caused mostly by the Compton (incoherent) scattering mechanism [16]. The absorption of X-rays within a material can easily be quantified using the Beer-Lambert law [17]. In case of kapton foil, we found a transmission > 98 % at 20 keV X-ray energy for foil thickness values ranging from 12 to 100 μm [18]. Even though there is a finite probability that some of the incident photons will scatter while interacting with kapton foil. As mentioned above contribution of Compton scattering plays significant role in altering the spectral background during the EDXRF measurements. Its contribution significantly influences the net area fluorescence intensity observed for an element and becomes a primary source of uncertainty during quantitative analysis. The intensity of the scattered radiation is significantly influenced not only by the energy of the primary X-ray beam, but also by the compositional properties of the specimen as well as sample carrier. X-rays emitted from a synchrotron source are highly linearly polarized in the plane parallel to electron orbiting plane. In such a case, it may be worth important to understand anisotropic nature of the emission profile of scattered radiation [19] in order to realize limitations for the placement of an x-ray spectroscopy detector in a condition where the contribution of the scattered background is minimal.

The differential Compton scattering cross-section and differential Rayleigh scattering cross-section for a polarized synchrotron photon of energy E, which scatters at finite polar angle $\phi$ (azimuthal angle) and at a specific scattering angle $\theta$, can be determined using the equations (1) and (2) respectively,

$$\frac{d\sigma_{Com}(\theta,\phi,E)}{d\Omega} = \frac{r_e^2}{2} \left(\frac{K}{K_0}\right)^2 \left(\frac{K}{K_0} + \frac{K_0}{K} - 2\sin^2\theta \cos^2\phi\right) S(x, Z) \quad (1)$$

$$\frac{d\sigma_{Ray}(\theta,\phi,E)}{d\Omega} = r_e^2 (1 - \sin^2\theta \sin^2\phi) F^2(x, Z) \quad (2)$$

$\frac{K_0}{K} = 1 + \frac{E_i}{mc^2}(1 - \cos\theta)$ represents the ratio of incident photon energy $E_i$ and scattered photon energy E. x equates $(E/hc) \sin(\theta/2)$, $r_e$ describes the classical radius of the electron. $S(x,Z)$ is an incoherent scattering function where as $F(x, Z)$ describes atomic form factor. Their details can be obtained from elsewhere [20, 21].

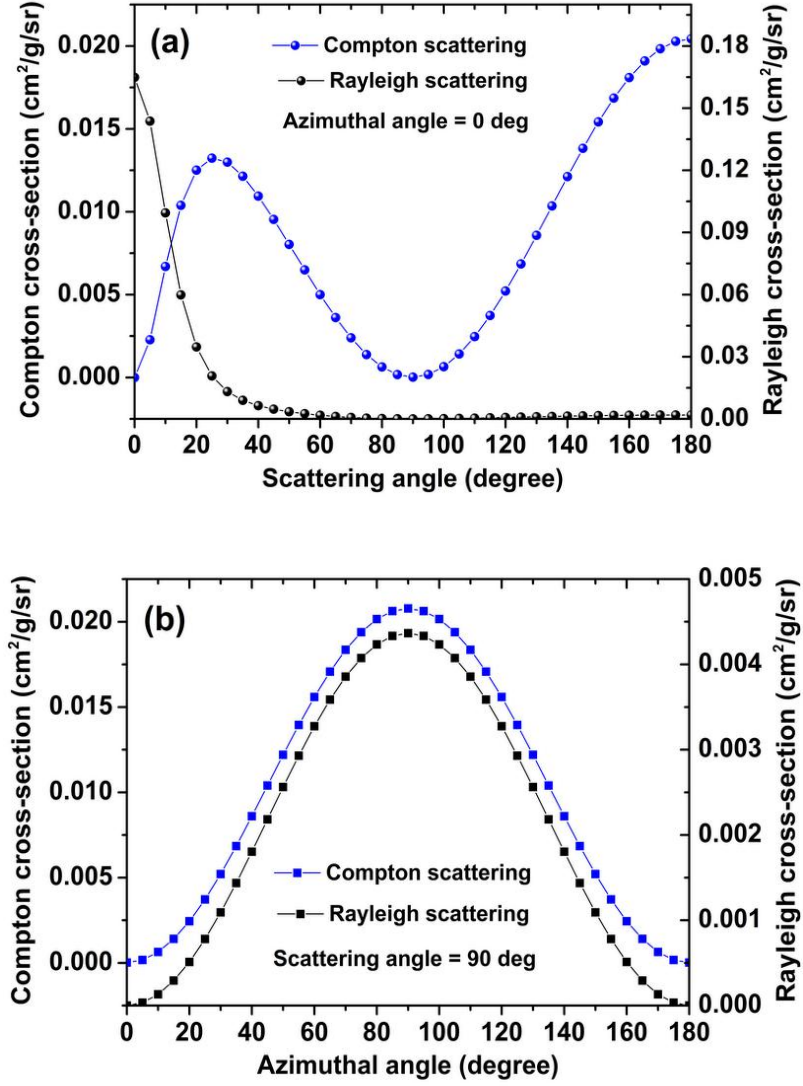

**Fig. 3.** Simulated results for angular distribution of scattered photons from kapton material. (a) Computed differential Compton and Rayleigh scattering cross-sections as a function of scattering angle θ at a fixed azimuthal angle of ϕ = 0° and, (b) variation of scattering cross section as a function of azimuthal angle ϕ at a fixed scattering angle θ = 90°.

Figs. 3 show variation of differential Compton scattering cross-section and Rayleigh scattering cross-section as a function of scattering and azimuthal angles in case of polarized synchrotron photons of energy 20 keV while they interact with kapton material. The computation were carried out using the xraylib database [22]. It can be seen from Fig. 3(a) that Compton scattering cross section varies in a nonlinear manner as a function of scattering angle whereas Rayleigh scattering cross-section shows maximum value at θ = 0° and it decreases rapidly upto θ = 40°. At larger scattering angles, variation in Rayleigh scattering cross-section is insignificant and its value remains more or less constant. The dependency of Compton scattering and Rayleigh scattering cross-sections (Fig.3b) on the polar azimuthal angle ( ϕ ) demonstrates their sinusoidal nature. One finds maximum value of the scattering cross sections at ϕ = 90° while it shows minima for ϕ = 0° and 180°, which fairly coincide with the S-polarization plane of synchrotron X-ray beam. This implies that if the spectroscopy detector is placed in the horizontal geometry (i.e. θ = 90° and ϕ = 0°), one will observe minimum spectral background in the florescence spectrum for both EDXRF and TXRF measurements. Above theoretical simulations provide us a basic guideline to setup a optimum experimental geometry for both EDXRF and TXRF measurements in order to achieve enhanced signal to scattered background ratio while employing a synchrotron x-ray beam as an excitation source.

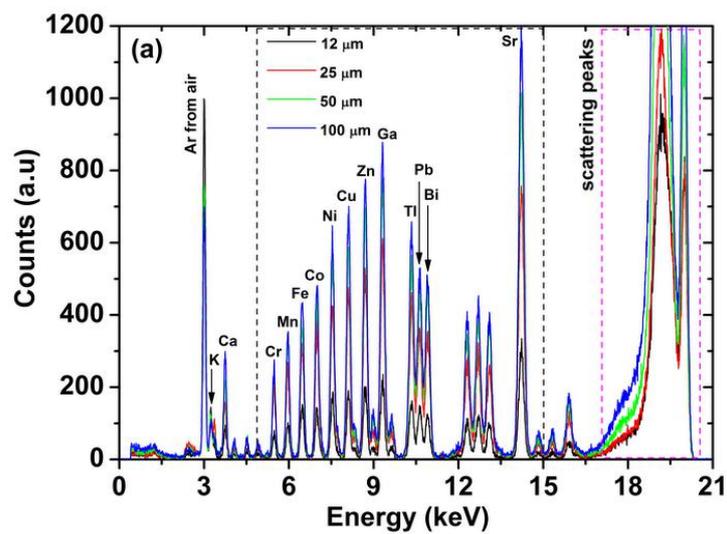

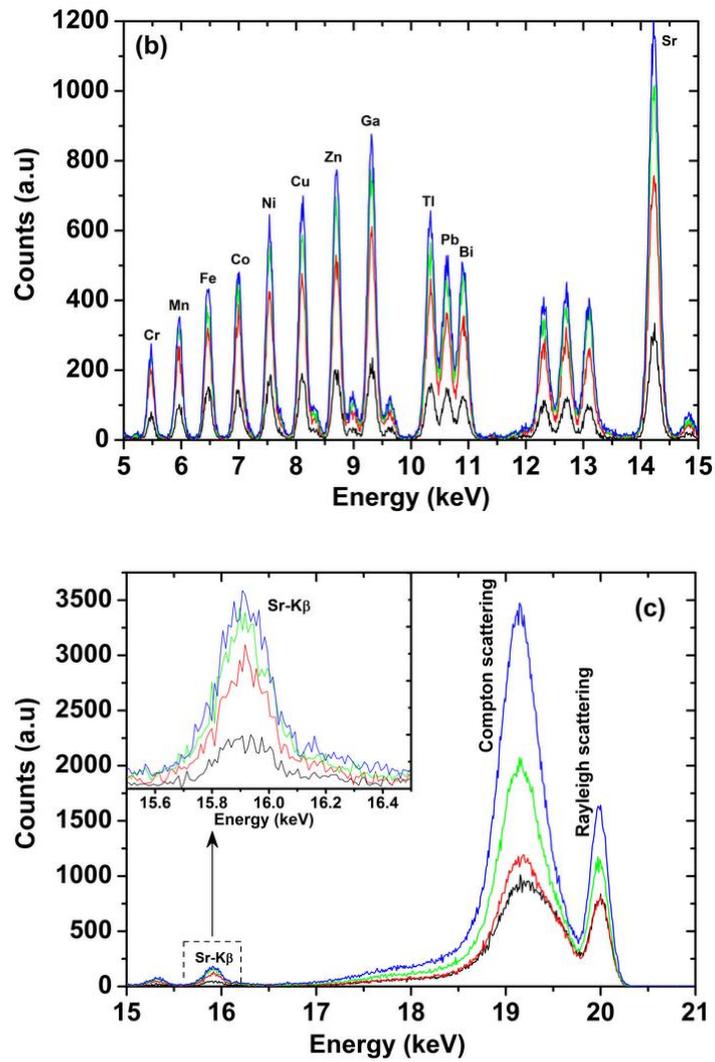

**Fig. 4.** XRF spectra obtained from a multi-element standard reference sample (ICP-IV) employing kapton foils with varying thicknesses as sample carrier. All the spectra were recorded at 20 keV excitation energy with a detector live time of 100 seconds in similar experimental conditions. (a) full spectrum showing all detected elements along with scattered peaks, (b) expanded view of observed fluorescence peaks for different elements, (c) expanded view of Compton and Rayleigh scattered peaks. The inset of Fig. (c) demonstrates the influence of tailing background of Compton scattered peak on Sr-Kβ fluorescence intensity.

**Measured results**

Figs. 4(a)-(c) show measured EDXRF spectra obtained from an ICP-IV standard reference material containing 90 ng of each element on top of kapton foils of different thickness values at monochromatic X-ray energy of 20 keV. The figure clearly confirms the presence of different elements in the ICP-IV sample of atomic number Z ranging from 19 (K) to 38 (Sr). The fluorescence peak of Ar was also observed, which is primarily generated from ambient air, as all measurements on the ICP-IV standard sample were carried out in the air ambient environment. Fig 4(b) depicts an expanded view of the measured spectra wherein fluorescence peaks of different elements have been shown for clarity. It can be noticed from Fig. 4(b), that the signal to background ratio of different detected elements unambiguously increases if one increases thickness of the kapton foil (i.e. sample carrier). The intensity gain in the fluorescence peak of an element saturates if thickness of kapton is taken between 50 to 100 $\mu$m. We observed similar kind of behavior in case of all detected elements. This dramatic change in the fluorescence intensity gain of an analyte element mainly arises due to enhanced scattered intensity produced from a kapton foil at larger thickness values. This indicates that, while scattered intensities (both Compton and Rayleigh) affect the spectral background in the XRF technique but they also contribute to the additional excitation of an analyte element in addition to the primary beam excitation. This feature may be particularly advantageous for detection of an analyte element that is present in trace amounts. Fig. 4(c), shows an enlarged view of a part of fluorescence spectrum where scattered peaks occur. It can be seen that intensities of both Compton and Rayleigh scattered peaks increase rapidly if thickness of the sample carrier (in our case kapton foil) is increased. The effect of scattered background on the intensity gain of a fluorescence peak becomes more noticeable if we observe Sr-Kβ peak (shown in the inset of Fig.4(c). These results indicate that spectral background has a considerable effect on the fluorescence intensity gain of an analyte element even when it is present at the tailing background of the Compton scattered peak.

In order to compare and evaluate the relative advantages of the kapton foil assisted EDXRF scheme with TXRF technique in terms of observed detection limits of different elements, we also carried out TXRF measurements on the ICP-IV multi-element standard by pipetting similar amount of analyte mass (~ 90 ng for each element) on a polished quartz glass substrate. Fig. 5 depicts measured TXRF spectrum with a detector acquisition time of 100 seconds. The detection limits of various elements in both cases (TXRF and EDXRF geometries) were estimated after determining net area intensity and observed spectral background under a fluorescence peak using PyMCA software [23, 24].

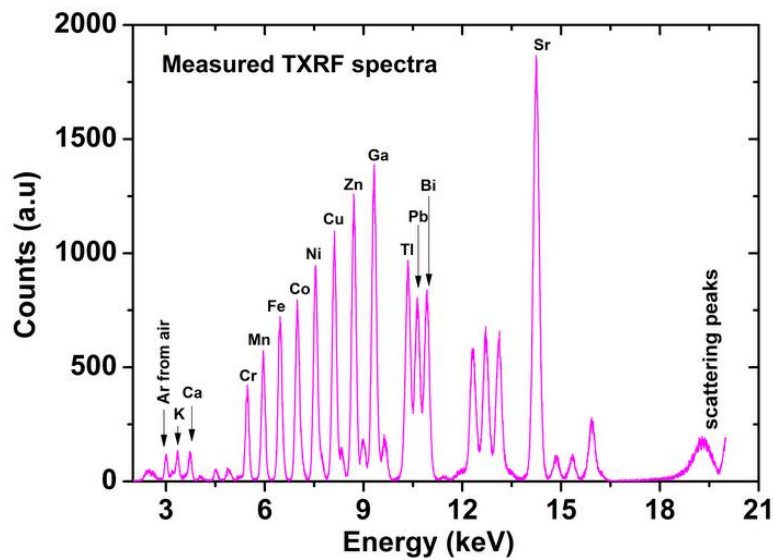

**Fig. 5.** TXRF spectrum obtained from an ICP-IV multielement standard containing ~ 90 ng mass of each element at excitation energy of 20 keV in the experimental geometry depicted in Fig.1(b) (θ = 90° and ϕ= 0°).

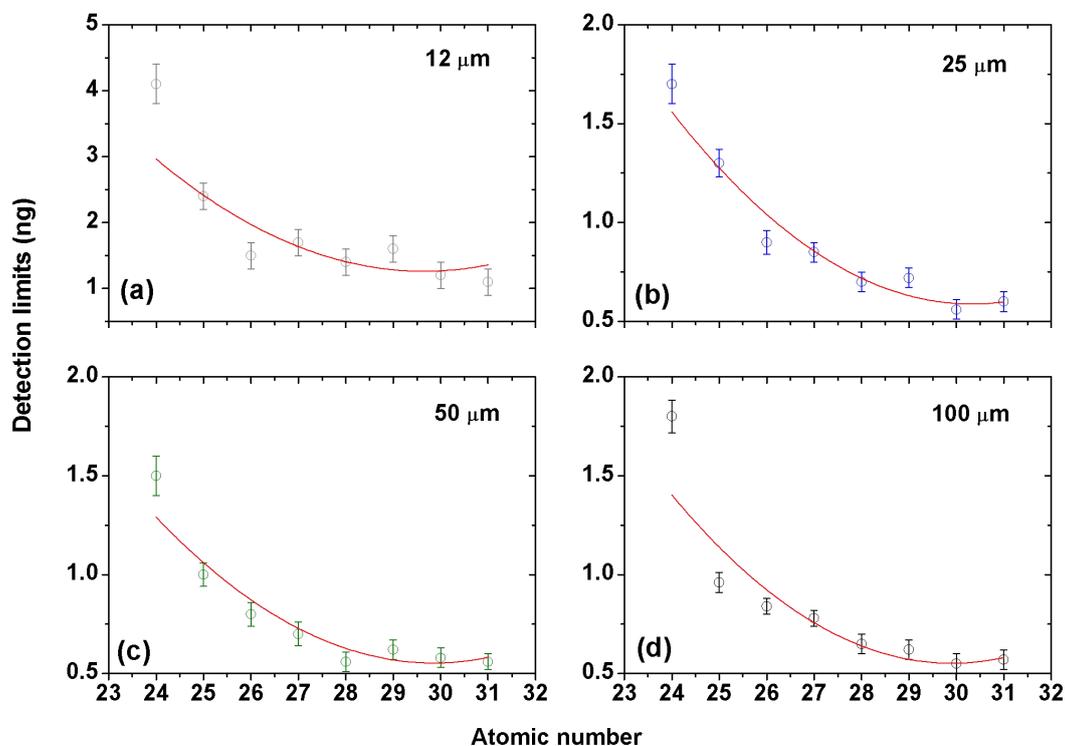

**Fig. 6.** Determined detection limits (DLs) for various elements using kapton foil as sample carrier with varying thickness values. (a) 12 μm, (b) 25 μm, (c) 50 μm and (d) 100 μm. In each case, ~ 90 ng mass of multi-element standard reference (ICP-IV) was used.

Figs. 6(a-d) illustrate variation of detection limits of different elements present in standard reference sample (ICP-IV), as a function of their atomic numbers, obtained using kapton foil assisted EDXRF measurements. Because the photoelectric absorption cross-section of an element nearly varies as $Z^4 E^{-3}$ (here Z represents atomic number of an element and E is the energy of incident X-ray beam) [17], one observes improved detection limit values when the absorption edge energy of an analyte

element is close to the excitation energy. As a result, the dependency of the detection limit on atomic number of an element produces a broad minima.

From Fig. 6, it can be clearly seen that one obtains improved detection limit values for majority of elements if kapton foil with thickness ranging from 25-50 micrometers are utilized as sample carrier during the EDXRF measurements. We have also noticed that lowering the thickness of kapton foil no longer improves the value of detection limit due to a reduction in analyte fluorescence yield. Because at lower thickness of kapton foil, the transmission of primary X-ray beam enhances which in turn reduces the scattered signal from the sample carrier, resulting in reduced detection limits for different elements. This effect was clearly realized in the case of 12 micrometer thick kapton foil (Fig. 6a). On the other hand at higher thickness values of kapton foil, DLs do not improve much rather remain more or less constant because the fluorescence intensity gain of an analyte element gets saturated, whereas the spectral background increases monotonously. These finding corroborate the experimental results depicted in Fig. 4 confirming the fact that an analyte element is excited not only by the primary x-ray beam but also by the scattered radiation generated from the sample carrier.

Fig. 7 illustrates detection limit values derived from the TXRF measurement for the same analyte (ICP-IV) containing 90 ng mass of each element on a polished quartz glass substrate.

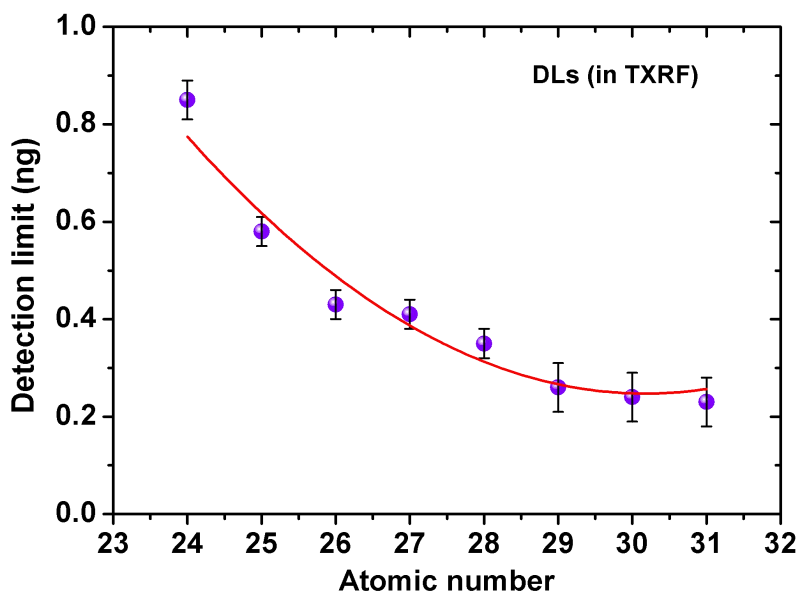

**Fig. 7.** Determined detection limits (DLs) as a function of atomic number using TXRF measurement. During TXRF investigation, similar (~ 90 ng) mass of the ICP-IV specimen was pipetted on top of a polished quartz glass substrate.

Table 1 summarizes determined detection limit values for different elements obtained using TXRF and kapton foil assisted EDXRF measurements (comprising of 25 μm thick kapton foil). To evaluate the benefits of kapton foil assisted EDXRF, we determined the ratio of DLs for each element. Tabulated data clearly show that relative strength of DL obtained using EDXRF is about 2 time poor (in magnitude) as compared to that of the TXRF method, which involves extremely complicated instrument alignment procedures.

Recently, it has been demonstrated that the elemental detection limits can also be increased by using a microfocused X-ray beam in conjunction with the thin kapton foil based EDXRF measurement geometry [25]. However, the spatial inhomogeneity of the sample residue deposited on kapton foil has a significant effect on the reliability of quantification procedures as well as the observed detection limits for different elements. On the other hand, conventional EDXRF ($45^0/45^0$ measurement geometry), allows full illumination of deposited sample residue, making it highly advantageous in addressing the aforementioned issues.

**Table 1.** Comparison of the detection limit values obtained using the EDXRF and TXRF techniques.

| Elements | Detection limits (ng) | | Ratios of DLs (EDXRF/TXRF) |
|---|---|---|---|
| | EDXRF | TXRF | |
| Cr | 1.69 ± 0.03 | 0.85 ± 0.04 | 1.99 |
| Mn | 1.30 ± 0.02 | 0.58 ± 0.03 | 2.24 |
| Fe | 0.93 ± 0.01 | 0.43 ± 0.03 | 2.16 |
| Co | 0.85 ± 0.01 | 0.41 ± 0.03 | 2.07 |
| Ni | 0.71 ± 0.01 | 0.35 ± 0.03 | 2.03 |
| Cu | 0.72 ± 0.02 | 0.26 ± 0.05 | 2.77 |
| Zn | 0.56 ± 0.02 | 0.24 ± 0.06 | 2.33 |
| Ga | 0.60 ± 0.01 | 0.23 ± 0.05 | 2.61 |

## CONCLUSIONS

On the basis of the systematic study presented here, we conclude that by using kapton foil as sample carrier with a thickness ranging from 25 to 50 μm, it is possible to obtain improved elemental detection limits while using the normal EDXRF method with a 45°- 45° measurement geometry. The proposed approach has been validated by analyzing a multi-element standard reference material comprising of several elements with comparable concentration values. It may be important to mention that the described approach greatly overcomes the limitations of the EDXRF technique, where rigorous corrections for matrix effects are often required during quantitative analysis of a thick pellet sample. Furthermore, because only a little mass of an analyte is used during the experimental measurement, the proposed method benefits from all the features of TXRF technique. We predict widespread applications of our method, particularly for the investigation of nuclear materials, because it considerably reduces the risk of radiation hazards for working personnel during measurements.

## ACKNOWLEDGMENTS

Authors acknowledge the technical staff of BL-16 beamline for their help in carrying out the experimental measurements. One of the authors Md. Akhlak Alam would like to thank RRCAT-HBNI for providing financial support.